\documentclass[aip,cha,amsmath,amssymb,superscriptaddress,floatfix]{revtex4-1}
\usepackage{siunitx}
\usepackage{gensymb}
\usepackage{graphicx}
\usepackage[caption=false]{subfig}
\begin{document}

\title{Balance of excitation and inhibition determines 1/f power spectrum in neuronal networks}
\author{F.Lombardi}
\affiliation{Institute of Computational Physics for Engineering Materials, 
ETH,Zurich,Switzerland}
\affiliation{Department of Industrial and Information Engineering, University of Campania ``Luigi Vanvitelli'', INFN Gr. Coll. Salerno, Aversa(CE), Italy}
\author{H.J.Herrmann}
\affiliation{Institute of Computational Physics for Engineering Materials, 
ETH,Zurich,Switzerland}
\affiliation{Departamento de F\'isica,Universitade Federal do Cear\'a, 
60451-970 Fortaleza,Cear\'a,Brazil}
\author{L.de Arcangelis}
\affiliation{Department of Industrial and Information Engineering,University of Campania ``Luigi Vanvitelli'', INFN Gr. Coll. Salerno, Aversa(CE), Italy}

\begin{abstract}

The $1/f$-like decay observed in the power spectrum of  electro-physiological signals, along with  scale-free statistics of the so-called neuronal avalanches,  constitute evidences of criticality in 
neuronal systems.   
Recent \textit{in vitro} studies have shown that avalanche dynamics at criticality corresponds to some specific  balance of excitation and inhibition, thus suggesting  that this is a basic feature of the critical state of neuronal networks. In particular,  a lack of inhibition significantly alters the temporal structure of the  spontaneous avalanche activity  and leads to an anomalous abundance of large avalanches.
Here we study  the relationship between network inhibition and the scaling exponent $\beta$ of the power spectral density (PSD) of avalanche activity in a neuronal network model inspired in Self-Organized Criticality (SOC). 
We find that this scaling exponent depends on the percentage of inhibitory synapses and  tends to the value $\beta = 1$ for a percentage of about 30\%. More specifically, $\beta$ is close to $2$, namely brownian noise, for purely excitatory networks and decreases towards values in the interval $[1,1.4]$  as the percentage  of inhibitory synapses ranges between 20 and 30\%, in agreement with experimental findings. 
These results indicate that the level of inhibition affects the frequency spectrum of resting brain activity and suggest the analysis of the PSD scaling behavior as a possible tool to study pathological conditions.    

\end{abstract}

\maketitle

\begin{quotation}
 
Power spectra of the form $S(f)\propto 1/f^\beta$, with $\beta \approx 1$, are the distinctive feature of processes with long-range temporal correlations and often appear in conjunction  with avalanche-like dynamics. In the case of neuronal systems,  neuronal avalanches show a power law size distribution with an exponent $\alpha \simeq -1.5$, which in turn is an evidence of long-range spatial correlations. Long-term spatio-temporal correlations are typical of systems at criticality.
Several studies performed on cultures of cortical neurons  indicate that criticality is connected to a specific ratio of inhibition and excitation. Indeed, in disinhibited networks the avalanche size distribution markedly deviates from a power law  and tends to exhibit a  bimodal functional form, with a higher probability for small and large avalanches.  
Here we investigate the influence of inhibitory synapses on the avalanche activity power spectrum by a numerical model inspired in SOC. This study aims to test, on a minimal neuronal network model, the idea  that  the power spectrum scaling behavior can give information on the  pathological conditions of neuronal systems.
\end{quotation}

\section{\label{sec:intro}Introduction}

\label{intro}
Cortical networks exhibit an intense spontaneous activity, defined as the activity of a neuronal population that is not driven by any  external stimuli or behavioral tasks. This property is common to all neuronal systems, regardless species, sizes and conditions. \textit{In vitro} as \textit{in vivo}, from leech ganglia to small patches of cortex in a physiological bath and all the way up to the entire human cortex, one does observe an intense endogenous activity, whose constant feature is the irregular sequence of periods of  synchronous firings.  

Spectral analysis is massively employed in the investigation of the brain and, more generally, neuronal networks activity. Beside the identification of oscillatory behavior, it plays a crucial role in  characterizing temporal correlations. Indeed the spectral density of a certain stochastic  process $x(t)$ is related to the correlation function $C(\tau)$ by the Wiener-Khinchin theorem, $S(f) \propto \int C(\tau)cos(2\pi \tau)d\tau$. A pure random process has no correlations in time and $S(f)=const$. Its integral is the well known Brownian motion characterized by  $S(f)\propto 1/f^2$, typical of a temporal process composed of  uncorrelated increments. On the other hand processes  with long-range temporal correlations are characterized by a power spectrum  $S(f)\propto 1/f^\beta$ with $\beta \approx 1$, the so-called $1/f$ noise. \\
Power laws in the Power Spectral Density (PSD) have been reported in many experimental studies involving EEG and MEG, as well as in the LFPs of ongoing cortical activity and in resting state fMRI \cite{novik97,deste06,deste10,pritchard92,zara97}. 
Local peaks usually observed in the PSD of these signals correspond to periodic oscillations and several procedures have been adopted in order to separate the oscillatory from the scale-free component \cite{novik97,yama93}.
In 1992 \textit{Prichard} examined the scaling behavior of the power spectrum of the human eyes-closed and eyes-open resting EEG \cite{pritchard92}. For the eyes-closed condition the power spectrum averaged over different brain areas showed a low frequency ($0.5-8$ Hz) power law decay with exponent $\beta = 1.32$, whereas in the eyes-open condition the exponent was $\beta = 1.27$. However for both  conditions $\beta$ was found to vary across brain regions and the standard error of estimate was in the interval $[0.26,0.28]$. A similar result has been recently obtained by \textit{Dehghani et al.}  \cite{deste10}, who have measured an average slope $\beta = 1.33\pm 0.19$ and again observed that the scaling of PSD changes across the brain, varying between $1/f$ in the midline channels and $1/f^2$ in the temporal and frontal ones \cite{deste10}. The same authors have found $\beta \simeq 1$ on the frequency range [$0.5, 20$Hz] of the PSD of bipolar LFPs recorded from cat parietal cortex during waking \cite{deste06}. 

The estimation of the scaling exponent of the PSD becomes more difficult when using MEG and results usually differ from the ones obtained from EEG. Indeed the detectors used in the  MEG recordings are very sensitive to environmental noise and can produce $1/f$ noise \cite{hama93}. Therefore, part of the scaling of the MEG power spectrum could be due to the filtering of the sensors and noise corrections are usually needed. For this reason results are more debated and less homogeneous. In an early study \textit{Novikov et al.} considered two kinds of signals: The normal component of the magnetic field from individual channels and the difference between the signals of two channels, thus subtracting alpha band peaks.
On the one hand, single channels exhibited a power law regime at low frequencies ([$0.5,8$Hz]) with exponent $\beta$ between 1.03 and 1.19 and a peak corresponding to the alpha frequency band. On the other hand, on the same frequency range ([$0.5,10$Hz]), the average exponent for the power spectrum of the difference was 0.98 in one subject and 1.28 in the other one.
The signal of individual channels has been recently investigated by \textit{Dehghani et al.} \cite{deste10}, who analyzed the scaling of both the MEG and EEG signal simultaneously in the range [$0.1,10$Hz] and found that the PSD of MEG signal averaged over all sensors exhibited an exponent $\beta = 1.24\pm 0.26$ before noise correction, and $\beta = 1.06\pm 0.29$ after. As for the EEG signal, they have  observed that the scaling of PSD changes across the brain: The exponent $\beta$ exhibits the highest value in the frontal area, it is close to 1 in the central area and takes the lowest value in a  parietotemporal horseshoe region.

Long-range temporal correlations and power spectra of the form $1/f^{\beta}$ have also been reported in the analysis of spontaneous brain oscillations. 
In particular, for the  amplitude modulation of $\alpha$ oscillations in MEG data, \textit{Linkenkaer-Hansen et al.} \cite{linken01} have found  $\beta=0.44\pm 0.09$ in the eyes-closed condition and $\beta=0.52\pm 0.12$ in the eyes-open condition. 
Similar exponents characterize the PSD scaling in the case of EEG recordings: $\beta=0.36\pm 0.17$ in the eyes-closed condition and $\beta=0.51\pm 0.12$ in the eyes-open condition \cite{linken01}. 

Finally we notice that evidences for a $1/f$ power law scaling of the PSD have been  found in the spontaneous fluctuations of fMRI signal \cite{zara97} as well. 
To summarize, we can state that experimental evidences are globally  consistent with long-range temporal correlations in ongoing healthy brain activity. Indeed the scaling exponent $\beta$ of the PSD at low frequencies always takes values within the interval $[0.8,1.5]$. On the other hand, for epileptic patients the exponent $\beta$ has been found to be in the range $[2.2,2.44]$ in the awake state, and $[1.6,2.87]$ in the Slow Wave Sleep \cite{he2014}.

According to Bak, Tang and Wiesenfeld (BTW) \cite{btw:soc}, the widespread occurrence of $1/f$ noise, together with self-similar spatial structures, indicate  that dynamical systems with extended spatial degrees of freedom self-organize into a critical state. In contrast with equilibrium critical phenomena, where a parameter has to be tuned to reach the critical point, these systems naturally evolve towards a state characterized by  long-range spatio-temporal correlations. BTW referred to this phenomenon as Self-Organized Criticality (SOC)\cite{btw:soc}.
The hypothesis that the brain, in particular the cortex, could exhibit critical behavior  was advanced following the argument that, to ensure an optimal functioning in a continuously changing environment, the brain should operate in a state which is at the border between order and disorder. The experimental discovery of neuronal avalanches  \cite{plenz:pl03} confirmed the SOC scenario. Neuronal avalanches are bursts of activity cascades  involving  a variable number of neurons and lasting up to hundred milliseconds \cite{plenz:pl03}.  The peculiar feature of neuronal avalanches is their scale-free statistics, namely  a power law behavior in the size and duration distributions. In particular, the power law scaling in the distribution of avalanche sizes is the fingerprint of long-range spatial correlations. Indeed, the existence of spatial correlations whose range is only limited by the system size \cite{plenz:pl03,shriki13}  constitutes an important experimental evidence for criticality \cite{pruessner}. 

In the following we study the power spectrum of avalanche activity in a neuronal network model inspired in SOC \cite{lda:06,fl2012,frontiers}. In particular, we investigate the relationship between the network inhibition and the scaling exponent of the avalanche activity power spectral density. 
We consider avalanche dynamics for different network topologies, i.e. regular square lattice, small world and scale-free, and consistently show that inhibitory neurons determine the scaling behavior of the PSD.     
Indeed, for a neuronal network of only excitatory synapses the PSD follows a power law with an exponent $\beta$ very close to $2$. By adding inhibitory neurons $\beta$ decreases and for a percentage  of inhibitory synapses between 20 and 30\%, $\beta$  exhibits values in the interval $[1,1.4]$, in agreement with experimental findings \cite{novik97,deste06,deste10,pritchard92,zara97}.

\section{Numerical model}
\label{model}
\subsection{Network structure}
We consider $N$ neurons as nodes of networks with three different topologies: Regular square lattice, small world and scale-free. In the first case neurons are sites of a square lattice of size $L \times L$ and are connected by directed synapses to their nearest neighbors. A directed synapse is such that it can be used by neuron $i$ to send a signal to neuron $j$ but not vice-versa. Each neuron $i$ has an out-going and in-going degree of connectivity $k_{out_i}=k_{in_i}=4$.\\  
Small world networks are obtained from the square lattice by rewiring one end of 10\% of the synapses to a neuron chosen at random in the network.
Finally, in the third case neurons are randomly distributed in a square and  connected by a scale-free network of synapses. More precisely to each neuron $i$ we assign an out-going connectivity degree, $k_{out_i}\in [2,100]$, according to the degree distribution $P(k)\propto k^{-2}$ of the functional network  measured in  \cite{chia:sfn} and two neurons are then connected with a distance-dependent probability, $P(r) \propto e^{-r/r_0}$, where $r$ is their Euclidean distance \cite{roerig02} and $r_0$ a typical edge length. 

To each synaptic connection we assign an initial random strength $g_{ij} \in [0.15, 0.3]$ and to each neuron an excitatory or inhibitory character. Outgoing synapses are excitatory if they belong to excitatory neurons, inhibitory otherwise.  Recent results \cite{bonifaz09} have shown that inhibitory neurons are hubs in scale-free functional networks. In our study we consider both cases, inhibitory neurons chosen at random and among the highly connected neurons. However, the different choices do not significantly affect the power spectrum data. 

Since synapses are directed, $g_{ij} \neq g_{ji}$, in general out-degree and in-degree of a neuron do  not coincide. Therefore once the network of output connections is established, we identify the resulting degree of in-connections, $k_{in_j}$, for each neuron $j$, namely we identify the number of synapses directed to  $j$. The number $k_{in_j}$ of in-going synapses can be considered as the dendritic tree of neuron $j$. We then assume that each neuron $j$ has a soma whose surface is proportional to $k_{in_j}$.

\subsection{Neuronal Dynamics}
Each neuron $i$ is characterized by a potential $v_i$, the so-called membrane potential, and fires if and only if its membrane potential $v_i$ is equal or above a certain threshold $v_{max}$, which we will refer to as firing threshold. In our simulations we set $v_{max} = 6$. However, as in every SOC-like model, this parameter is not relevant and results are independent of this particular choice. 
To trigger activity, we apply  to a random neuron a stimulation whose intensity is chosen at random in the interval $[v_{max}/6,v_{max}/3]$. Then, whenever at time $t$ the potential of neuron $i$ fulfills the condition $v_i \geq v_{max}$, the neuron fires and its potential $v_i$ arrives at each of the $k_{out_i}$ connected neurons.\\  
For real neurons the production of neurotransmitters at the presynaptic terminals \footnote{Given two neurons, $i$ and $j$, and a  synaptic connection directed from $i$ to $j$, $i$ is called presynaptic and $j$ postsynaptic neuron.} is controlled by the frequency of action potentials, which depends on the integrated stimulation  received by the neuron. Here the integrated stimulation is given by $v_i$, the membrane potential of the firing neuron, which stimulates equally all connected neurons. 
The change in the  membrane potential of the postsynaptic neuron $j$ due to the firing of neuron $i$ is proportional to the relative synaptic strength $g_{ji}/\sum_l g_{li}$, 
\begin{equation}
v_j(t+1)=\ v_j(t) \pm \frac{v_i\cdot k_{out_i}}{k_{in_j}} \frac{g_{ji}}{\sum_{l=1} ^{k_{out_i}} g_{li}}.
 \label{eqn:dv}
\end{equation}
The normalization of the synaptic strength ensures that during the propagation of very large avalanches the voltage potential assumes finite values, while plus or minus sign is for excitatory or inhibitory synapses, respectively. After firing, the membrane potential of the neuron is set to $v_{rest} = 0$ and the neuron  remains in a refractory state for $t_{ref} = 1$ time step, during which it is unable to receive or transmit any signal. Each neuron in the network is an integrate and fire unit, therefore it will change its potential by summing the stimulations of successive firing neurons according  to Eq. \ref{eqn:dv} for all network types. In the case of the square lattice, since $k_{out} = k_{in}$, Eq. \ref{eqn:dv}  reduces to 
\begin{equation}
v_j(t+1)=\ v_j(t) \pm v_i \frac{g_{ji}}{\sum_{l=1} ^{nn} g_{li}}, 
\label{eqn:dv_sql}
\end{equation}
where the sum runs over the nearest neighbors (nn) of neuron $j$.

\subsection{Synaptic plasticity}
When a neuron $i$ fires, its out-going connections induce a potential variation in the $k_{out_i}$ postsynaptic neurons. 
We then say that these synapses are active and their   
strength $g_{ji}$ is increased proportionally to the membrane potential variation $|\delta v_j|$ of the postsynaptic neuron $j$,
\begin{equation}
 g_{ji}=\  g_{ji} + | \delta v_j |/v_{max}.
 \label{eqn:spla}
\end{equation}                                    
Conversely, the strength of all inactive synapses is reduced by the average strength increase per bond
\begin{equation}
\Delta g=\sum_{ij} \delta g_{ji}/N_B,  
 \label{eqn:dg}
\end{equation}
where $N_B$ is the number of bonds.
We set a minimum and a maximum value for the synaptic strength $g_{ij}$, $g_{min}=0.0001$ and $g_{max}=1.0$. Whenever $g_{ij} < g_{min}$, the synapse $g_{ij}$ is pruned, i.e. permanently removed.
Eq. \ref{eqn:dg} implements a sort of homeostatic mechanism that dynamically balances synaptic strengthening in the network. 
Indeed  the larger the average synaptic strengthening,  the more inactive synapses are weakened. 
These rules constitute a Hebbian-like scheme for synaptic plastic adaptation. The network memorizes the most used synaptic paths by increasing their strengths, whereas less used synapses eventually are pruned.  

\subsection{Avalanche activity}
When the membrane potential of all neurons is lower than $v_{max}$, the random stimulation can make one of them fire and thus keep the system active.
The firing neuron may indeed bring to threshold some of the connected  neurons,  generating an avalanche, a cascade of  activity which propagates through the network involving a variable number of neurons. 
During an avalanche no further random stimulation is applied, as in usual SOC models. This procedure implements the separation of time scales between fast avalanche propagation and slow  avalanche triggering. In biological terms, we suppose that the avalanche propagation is due to successive activations of connected neurons and  neglect the possible contribution of spontaneous neurotransmitter release. 
As soon as no more neurons are able to fire, the avalanche ends and its  size is recorded as the number of firing neurons $s$, or, alternatively, as the sum $s_{\Delta V}$ of all positive potential variations (depolarizations)  $\delta v_i ^{+}$ occurred in the network, namely  $s_{\Delta V} = \sum_i \delta v_i ^{+}$. By definition a single neuron firing does not constitute an avalanche. Avalanches are also characterized by their duration $T$, which is defined as the number of iterations taken by the activity propagation. 
The numerical time step for each iteration corresponds to the real time between the triggering of an action potential in the presynaptic neuron and the change of the membrane potential in the postsynaptic neuron, therefore it is of the order of 4-6 ms. After an avalanche ends, external stimuli trigger further activity in the system. 

At the end of each avalanche we implement the plasticity rules defined above. Since cortical plasticity such as long-term potentiation acts on time scales of seconds to minutes, which is much longer than the duration of avalanches, we do apply the plasticity protocol for a number of stimulations and then study avalanche activity without further changing synaptic strengths. 
In particular, for scale-free networks, since we don't want to alter the connectivity degree distribution of the initial network, we apply plasticity rules until the first few synapses are pruned.

\begin{figure}[t!]
\begin{center}
\includegraphics[width=14.5cm]{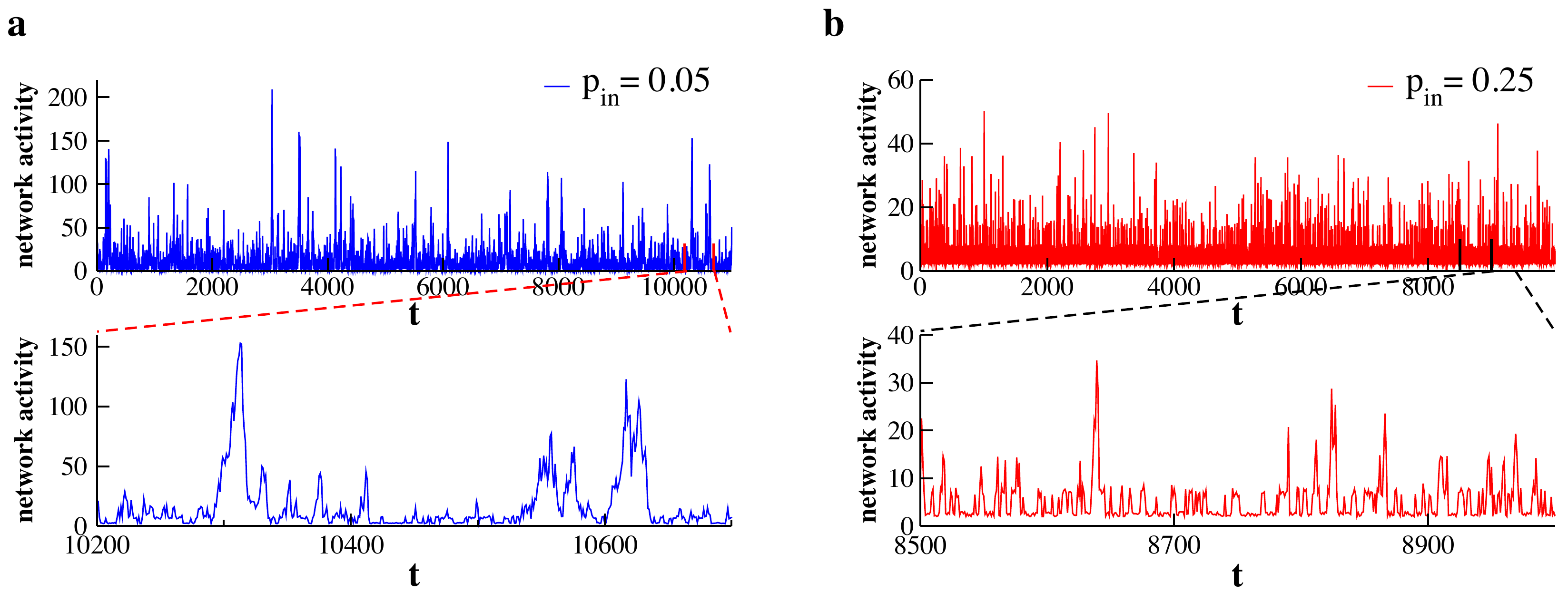}
\end{center}
\caption{Network activity at different time scales and for different fraction $p_{in}$ of inhibitory synapses. The intensity of activity $V(t)$ is the sum of all potential variations occurring in the network at each time step, namely $V(t) = \sum_i \delta v_i (t)$. a) $p_{in}=0.05$; b)$p_{in}=0.25$.}
\label{fig:signal}
\end{figure}

For each network configuration we construct a temporal signal $V(t) = \sum_i \delta v_i (t)$ (Fig. \ref{fig:signal}) as the sum  of all potential variations occurring at each time step during the network activity and interpret it as the 'EEG' of the system. We then consider the  PSD of this signal as the PSD of network activity. 

\section{Results}

\begin{figure}[t!]
\begin{center}
\includegraphics[width=16.8cm]{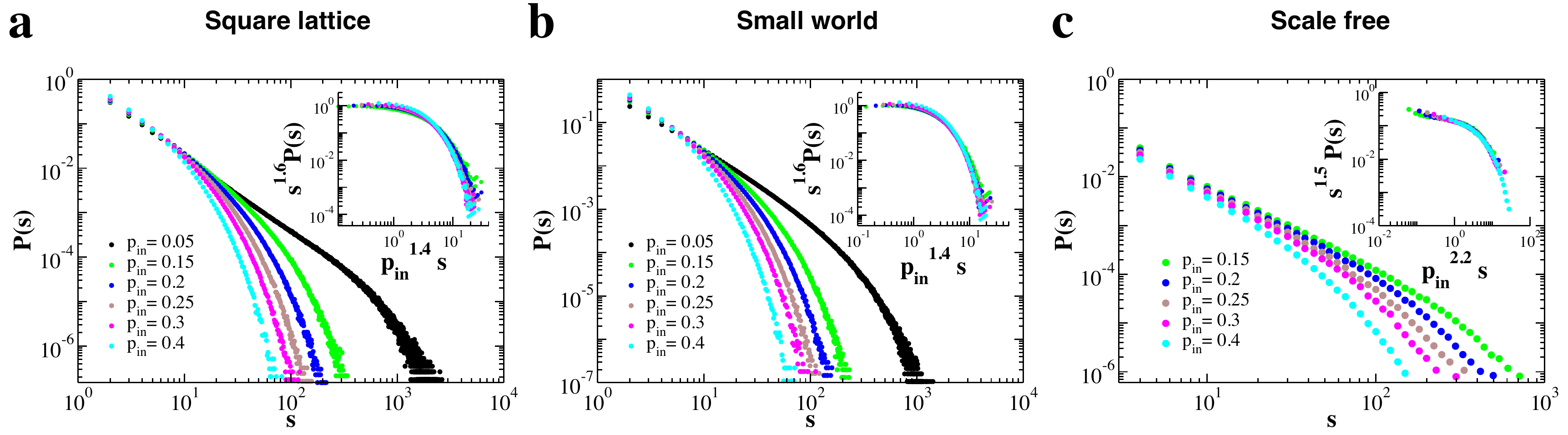}
\end{center}
\caption{Avalanche size distributions for different values of $p_{in}$. a) Regular square lattices with $N=10000$ neurons. Inset: Plotting $ s^{\alpha} P(s)$ vs $p_{in} ^{\theta} s$, with $\alpha = 1.6$ and $\theta = 1.4$, data collapse onto a universal scaling function. b) Small world networks  with $N=10000$ neurons. Inset: Plotting $ §s^{\alpha} P(s)$ vs $p_{in} ^{\theta} s$, with $\alpha = 1.6$ and $\theta = 1.4$, data collapse onto a universal scaling function. c) Scale-free network with $N=64000$. Inset: Plotting $ s^{\alpha} P(s)$ vs $p_{in} ^{\theta} s$, with $\alpha = 1.5$ and $\theta = 2.2$, data collapse onto a universal scaling function. 
}
\label{fig:size_sql}
\end{figure}

To understand the role of inhibitory neurons in avalanche activity, we first analyze the distribution $P(s)$ of avalanche sizes for different fractions  $p_{in}$ of inhibitory synapses and the three different network topologies (Fig. \ref{fig:size_sql}). For very small values of $p_{in}$ we observe a power law regime over about two decades followed by an exponential cutoff.  By increasing the percentage of inhibitory synapses  the exponential cutoff gradually moves towards smaller $s$ values and the scaling  regime  reduces, essentially disappearing for $p_{in} \ge 0.3$ (Fig. \ref{fig:size_sql}). 
This behavior can be expressed  by the following universal scaling relation
\begin{equation}
P(s) \sim s^{-\alpha} f(s/p_{in} ^{-\theta}).
\label{s_scaling}
\end{equation}
where $f(s/p_{in} ^{-\theta})$ is a scaling function.
Eq.5 contains two parameters: the scaling exponent $\alpha$ and the cutoff exponent $\theta$ expressing the dependence of the cutoff on $p_{in}$. The combination of the two parameters provides an apparently variable exponent for the distribution, as shown in the main Fig. \ref{fig:size_sql}. By plotting $P(s) s^{\alpha}$ vs. $s/p_{in} ^{-\theta}$, data should collapse onto the universal function $f$ for the correct values of $\alpha$ and $\theta$. Therefore searching for the collapse of the curves will provide the correct exponent values. 
As shown in the insets of Fig. \ref{fig:size_sql}, the  size distributions for different values of $p_{in}$ collapse onto a unique function for the following exponent values:
$\alpha = 1.6$ and $\theta = 1.4$ for the square lattice and the small world  networks  (Fig. \ref{fig:size_sql}a,b);  $\alpha = 1.5$ and $\theta = 2.2$ for scale-free networks  (Fig. \ref{fig:size_sql}c).

The behavior observed in Fig. \ref{fig:size_sql} is due to the dissipative role played by inhibitory synapses in the model  and is in agreement with previous results on dissipative SOC model \cite{rios:99}. Indeed, if a neuron $j$ is connected to an inhibitory neuron $i$, its potential $v_j$ decreases whenever neuron  $i$ fires, moving away from the firing threshold $v_{max}$. As a consequence, when an avalanche reaches inhibitory neurons, it dies out and, as  the number of those neurons increases, the probability that  avalanches propagate throughout the entire system consistently decreases. 

\begin{figure}[t!]
\begin{center}
\includegraphics[width=13.6cm]{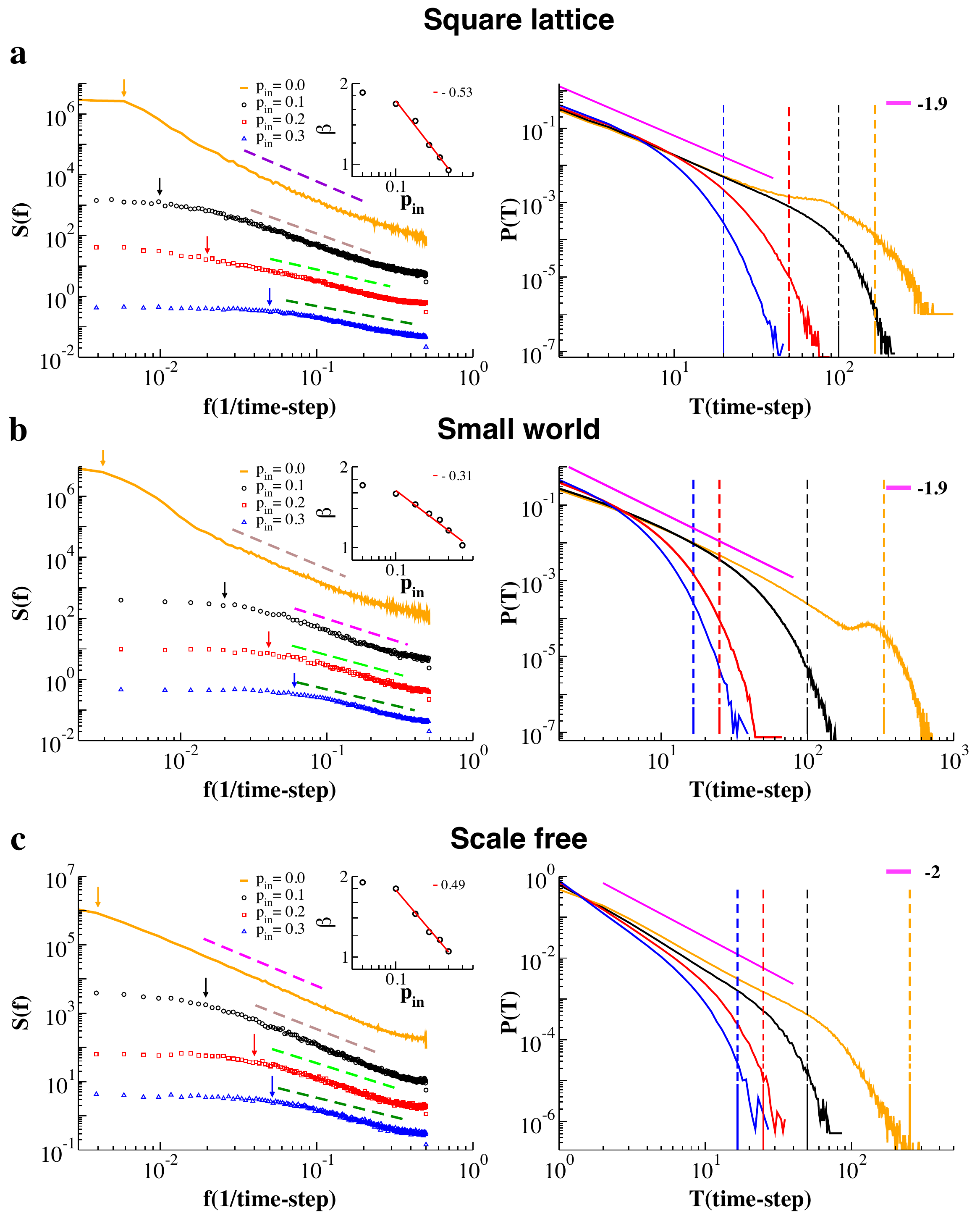}
\end{center}
\caption{Power spectral density (PSD) of avalanche activity (left column) and duration distribution $P(T)$ (right column) for different $p_{in}$. The PSD follows a power law whose exponent $\beta$ depends on $p_{in}$ and approaches the value $\beta = 1$  for $p_{in}=0.3$. The fitting intervals are indicated with arrows. The cutoff at low frequencies (arrows), which indicates the transition to white noise, scales with $p_{in}$ and corresponds to avalanche durations in the exponential cutoff of $P(T)$ (dashed lines).  a) Square lattice with $N=10000$ neurons.  b) Small world network with $N=10000$ neurons. c) Scale-free network with $N=64000$ neurons. Insets: Log-log plot of the exponent $\beta$ as a function of $p_{in}$.}
\label{fig:ps_pinhi_sql}
\end{figure}

This property of inhibitory synapses turns out to have a strong influence on  the PSD of avalanche activity $S(f)$. In Fig. \ref{fig:ps_pinhi_sql} we  show  $S(f)$ as a function of $p_{in}$.  
In our simulations the time unit is the time between the firing of the presynaptic neuron and the voltage change in the postsynaptic neuron which, in real systems, should correspond to few ms. With this rough correspondence the PSD frequency range is between about 1 and 100 Hz. The PSD has the same qualitative behavior for each value of $p_{in}$, namely an $f^{-\beta}$ decay and a cutoff at low frequencies that, as the cutoff in the avalanche size distribution shifts towards smaller $s$ values, moves towards higher frequencies with increasing $p_{in}$. The low frequency cutoff indicates the transition to white noise, which characterizes uncorrelated process, and  corresponds to avalanche durations in the exponential cutoff of the distribution $P(T)$ of avalanche durations (Fig. \ref{fig:ps_pinhi_sql}), as previously reported for a critical branching model \cite{poil08}. 
At the same time the exponent $\beta$ decreases for increasing values of the fraction of inhibitory synapses (Fig. \ref{fig:ps_pinhi_sql}).

\begin{figure}[b!]
\begin{center}
\includegraphics[width=9cm]{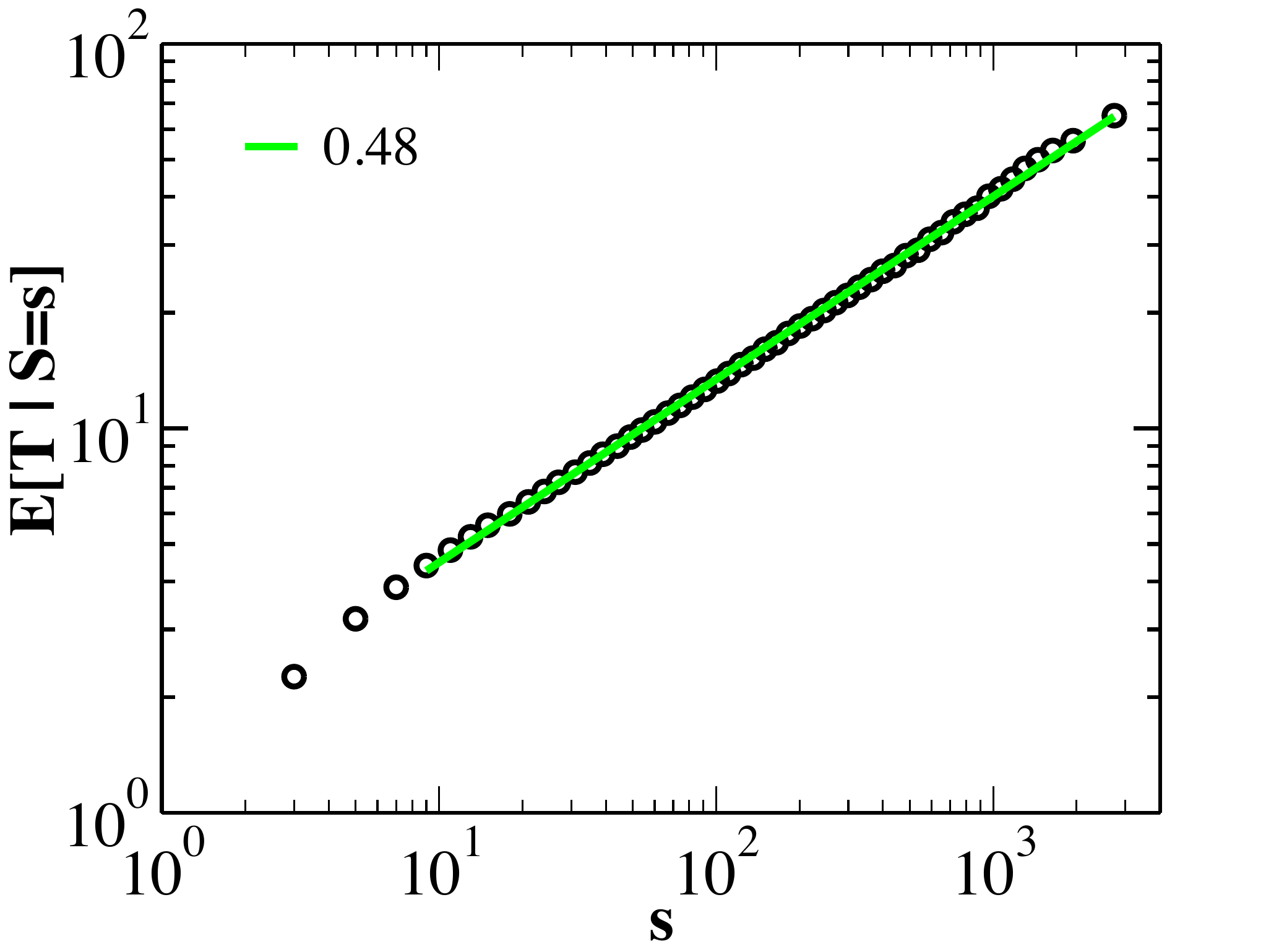}
\end{center}
\caption{Expectation value of avalanche duration $T$ for a given avalanche size $s$ on a scale-free network with $N = 4000$ and $p_{in}=0.05$: $E(T,s) \sim s^{\sigma \nu z}$, with $\sigma \nu z = 0.48$. Similar results are obtained for the square lattice and the small world network.}
\label{st}
\end{figure}

In particular, for a purely excitatory network we find that $\beta$ is  close to 2, an exponent associated to the PSD of the Brownian motion and larger than values found in experimental studies of spontaneous cortical activity in healthy subjects. Indeed also the duration distributions $P(T)$ for purely excitatory networks show evidence for very long, ``supercritical'' avalanches in regular and small-world networks (Fig. \ref{fig:ps_pinhi_sql}). However, when $p_{in}$ becomes closer to the fraction of inhibitory  synapses measured in real neuronal networks, i.e. about $0.3$, we find that $\beta$ is in  the interval $[1,1.4]$, the range of  experimentally measured $\beta$ values \cite{novik97,deste06,deste10,pritchard92,zara97}.  
More specifically, for $0.1 \le p_{in} \le 0.3$ the exponent $\beta$ decays as $p_{in} ^{-\delta}$, where $\delta \in [0.36,0.48]$, and tends to $1$ as $p_{in}\to 0.3$. \\
The scaling exponent $\beta$ of the PSD is related to the critical exponent $1/\sigma \nu z$, which  connects avalanche sizes and durations \cite{jsn:soc}, $s(T)\sim T^{1/\sigma \nu}$. It has been shown that, for purely excitatory models with $\alpha < 2$, $\beta = 1/\sigma \nu z$ \cite{sethna00}. Our model very closely follows this analytical prediction: Indeed for a purely excitatory network we find $\beta \simeq 2$ and $1/\sigma \nu z \simeq 2$, as shown in Fig. \ref{fig:ps_pinhi_sql} and \ref{st}. 
On the other hand, to our knowledge no analytical derivation of the relationship between $\beta$ and $1/\sigma \nu z$ is available for systems with excitatory and inhibitory interactions. Deriving such a relationship for SOC-like models with inhibitory interactions is a general problem of great interest and implies the introduction of anti-ferromagnetic interactions in the model of ref. \cite{sethna00}. This investigation is not a trivial extension of previous results  and  will be the subject of future studies.  

\begin{figure}[t!]
\begin{center}
\includegraphics[width=16.8cm]{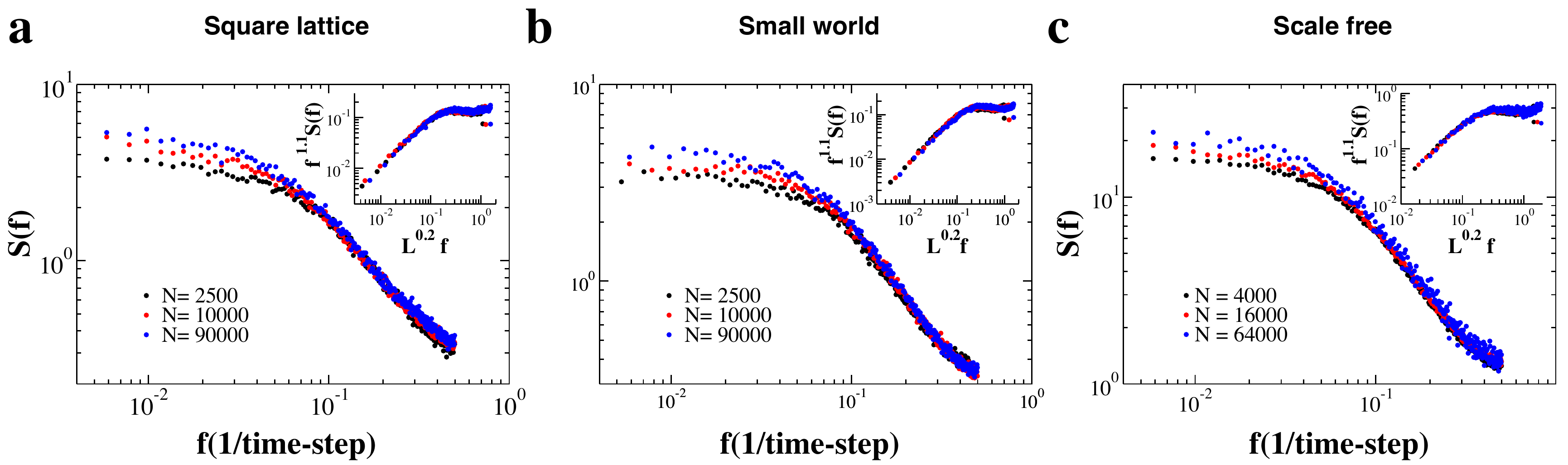}
\end{center}
\caption{The PSDs for different network sizes $N$ and $p_{in}=0.3$. a) Square lattice. b) Small world network. c) Scale-free network. Insets: Data  collapse onto a universal  function, according to the following scaling behavior, $S(f) \sim  f^{-\beta} g(f/L^{-\gamma})$, with $\beta = 1.1$ and $\gamma = 0.2$}
\label{fig:ps_pinhi03}
\end{figure}

Next we fix $p_{in}=0.3$ and investigate how $S(f)$ depends on  the system size $L$. We find that the low frequency cutoff shifts towards lower frequencies with increasing system size (Fig. 4) and the PSD follows the universal scaling behavior
\begin{equation}
S(f) \sim  f^{-\beta} g(f/L^{-\gamma})
\end{equation}
with $\beta = 1.1$ and $\gamma = 0.2$, as shown in the insets of Fig. \ref{fig:ps_pinhi03} . The values
of these exponents are not found to vary with the topology of the network.

\section{Discussion}

We have shown that a minimal neural network model inspired in SOC is able to capture one of the most intriguing statistical features of real neuronal networks, namely the $1/f$ decay of their ongoing activity PSD, $S(f)$ \cite{novik97,deste06,deste10,pritchard92,zara97}. 
We tested the model on three different network architectures and consistently found similar results: The exponent $\beta$, which characterizes the scaling behavior of $S(f)$, is  function of the fraction $p_{in}$ of inhibitory synapses  and tends to the value $\beta = 1$ for $p_{in} = 0.3$. Moreover, results are robust with respect to changes in the model parameters.

This $1/f$ decay in the frequency spectrum is the distinctive characteristics of processes with long-range temporal correlations and  often appears in conjunction with  avalanche-like dynamics. In the case of neuronal systems, avalanches show a power law size distribution with an exponent $-1.5$, which in turn is an indication of long-range spatial correlations \cite{plenz:pl03,shriki13}.
According to BTW, these long-term spatio-temporal correlations are typical of systems that self-organize into a critical state,  
whose paradigmatic model is the so called 'sandpile' model\cite{btw:soc}. 
This model exhibits a scale-free distribution for avalanche size \cite{btw:soc} and a power spectrum of the avalanche signal of the form $1/f^\beta$, but neither  $\beta$ is close to $1$ nor the exponent of size distribution is $-1.5$.
Several variants of the original sandpile model have been proposed in order to obtain an exponent $\beta \simeq 1$ \cite{rios:99,davidsen02,baie:sand}. In particular, \textit{de los Rios} et al. \cite{rios:99} have shown that a dissipative term in the dynamics of the original sandpile  gives rise to avalanche activity whose power spectrum is $1/f$.  However, in their model the avalanche sizes are not distributed according to a power law.\\
As pointed out above, in our model, inhibitory synapses play a dissipative role and limit  avalanche sizes. Nevertheless, as the data collapse shows, they  do not change the functional form of the size  distribution, which is a power law with an exponential cutoff. At the same time, the dissipation due to inhibitory synapses is crucial to get $S(f) \simeq 1/f$, as reported by \textit{de los Rios} et al  \cite{rios:99}. 
Indeed  the scaling behavior of the PSD, in the model is controlled by the  ratio  between the number of excitatory and inhibitory synapses, that is the ratio between excitation and inhibition.

The balance between excitation and inhibition plays a crucial role in the normal, non-pathological functioning of neuronal networks.
For instance, it has been shown that altering the balance between excitation and inhibition causes major changes in their spontaneous activity. In particular, blocking inhibition, or enhancing excitation, severely affects the power law behavior of avalanche size distribution and alters the dynamic range of the network \cite{plenz:pl03,shew09,copelli06}. More specifically, while in the normal condition this distribution follows a power law with an exponent $-1.5$, in disinhibited cultures the distribution  is bimodal, showing a higher   likelihood for small and large avalanches \cite{shew09}. 
In addition, the relationship between quiescence and activity at criticality appears to depend on the balance excitation-inhibition \cite{fl2016}. Indeed, it is significantly altered by suppressing inhibition, implying that inhibition plays an important role in the temporal organization of spontaneous activity \cite{fl2016,fl2012}.

Results presented here indicate that a deficit of inhibition may alter the frequency spectrum of resting brain activity and suggest the analysis of the PSD scaling behavior as a possible tool to investigate pathological conditions.

\newpage

\bibliography{bib_chaos_aip}

\begin{thebibliography}{29}%
\makeatletter
\providecommand \@ifxundefined [1]{%
 \@ifx{#1\undefined}
}%
\providecommand \@ifnum [1]{%
 \ifnum #1\expandafter \@firstoftwo
 \else \expandafter \@secondoftwo
 \fi
}%
\providecommand \@ifx [1]{%
 \ifx #1\expandafter \@firstoftwo
 \else \expandafter \@secondoftwo
 \fi
}%
\providecommand \natexlab [1]{#1}%
\providecommand \enquote  [1]{``#1''}%
\providecommand \bibnamefont  [1]{#1}%
\providecommand \bibfnamefont [1]{#1}%
\providecommand \citenamefont [1]{#1}%
\providecommand \href@noop [0]{\@secondoftwo}%
\providecommand \href [0]{\begingroup \@sanitize@url \@href}%
\providecommand \@href[1]{\@@startlink{#1}\@@href}%
\providecommand \@@href[1]{\endgroup#1\@@endlink}%
\providecommand \@sanitize@url [0]{\catcode `\\12\catcode `\$12\catcode
  `\&12\catcode `\#12\catcode `\^12\catcode `\_12\catcode `\%12\relax}%
\providecommand \@@startlink[1]{}%
\providecommand \@@endlink[0]{}%
\providecommand \url  [0]{\begingroup\@sanitize@url \@url }%
\providecommand \@url [1]{\endgroup\@href {#1}{\urlprefix }}%
\providecommand \urlprefix  [0]{URL }%
\providecommand \Eprint [0]{\href }%
\providecommand \doibase [0]{http://dx.doi.org/}%
\providecommand \selectlanguage [0]{\@gobble}%
\providecommand \bibinfo  [0]{\@secondoftwo}%
\providecommand \bibfield  [0]{\@secondoftwo}%
\providecommand \translation [1]{[#1]}%
\providecommand \BibitemOpen [0]{}%
\providecommand \bibitemStop [0]{}%
\providecommand \bibitemNoStop [0]{.\EOS\space}%
\providecommand \EOS [0]{\spacefactor3000\relax}%
\providecommand \BibitemShut  [1]{\csname bibitem#1\endcsname}%
\let\auto@bib@innerbib\@empty
\bibitem [{\citenamefont {Novikov}\ \emph {et~al.}(1997)\citenamefont
  {Novikov}, \citenamefont {Novikov}, \citenamefont {Shannahoff-Khalsa},
  \citenamefont {Schwartz},\ and\ \citenamefont {Wright}}]{novik97}%
  \BibitemOpen
  \bibfield  {author} {\bibinfo {author} {\bibfnamefont {E.}~\bibnamefont
  {Novikov}}, \bibinfo {author} {\bibfnamefont {A.}~\bibnamefont {Novikov}},
  \bibinfo {author} {\bibfnamefont {D.}~\bibnamefont {Shannahoff-Khalsa}},
  \bibinfo {author} {\bibfnamefont {B.}~\bibnamefont {Schwartz}}, \ and\
  \bibinfo {author} {\bibfnamefont {J.}~\bibnamefont {Wright}},\ }\bibfield
  {title} {\enquote {\bibinfo {title} {Scale-similar activity in the brain},}\
  }\href@noop {} {\bibfield  {journal} {\bibinfo  {journal} {Phys.Rev. E}\
  }\textbf {\bibinfo {volume} {56}},\ \bibinfo {pages} {R2387} (\bibinfo {year}
  {1997})}\BibitemShut {NoStop}%
\bibitem [{\citenamefont {Bedard}, \citenamefont {Kr\"oger},\ and\
  \citenamefont {Destexhe}(2006)}]{deste06}%
  \BibitemOpen
  \bibfield  {author} {\bibinfo {author} {\bibfnamefont {C.}~\bibnamefont
  {Bedard}}, \bibinfo {author} {\bibfnamefont {H.}~\bibnamefont {Kr\"oger}}, \
  and\ \bibinfo {author} {\bibfnamefont {A.}~\bibnamefont {Destexhe}},\
  }\bibfield  {title} {\enquote {\bibinfo {title} {Does the 1/f frequency
  scaling of brain signals reflect self-organized critical states?}}\
  }\href@noop {} {\bibfield  {journal} {\bibinfo  {journal} {Phys.Rev.Lett.}\
  }\textbf {\bibinfo {volume} {97}},\ \bibinfo {pages} {118102} (\bibinfo
  {year} {2006})}\BibitemShut {NoStop}%
\bibitem [{\citenamefont {Dehghani}\ \emph {et~al.}(2010)\citenamefont
  {Dehghani}, \citenamefont {Bedard}, \citenamefont {Cash}, \citenamefont
  {Halgren},\ and\ \citenamefont {Destexhe}}]{deste10}%
  \BibitemOpen
  \bibfield  {author} {\bibinfo {author} {\bibfnamefont {N.}~\bibnamefont
  {Dehghani}}, \bibinfo {author} {\bibfnamefont {C.}~\bibnamefont {Bedard}},
  \bibinfo {author} {\bibfnamefont {S.~S.}\ \bibnamefont {Cash}}, \bibinfo
  {author} {\bibfnamefont {E.}~\bibnamefont {Halgren}}, \ and\ \bibinfo
  {author} {\bibfnamefont {A.}~\bibnamefont {Destexhe}},\ }\bibfield  {title}
  {\enquote {\bibinfo {title} {Comparative power spectral analysis of
  simultaneous elecroencephalographic and magnetoencephalographic recordings in
  humans},}\ }\href@noop {} {\bibfield  {journal} {\bibinfo  {journal} {J.
  Comput. Neurosci.}\ }\textbf {\bibinfo {volume} {21}},\ \bibinfo {pages}
  {405--421} (\bibinfo {year} {2010})}\BibitemShut {NoStop}%
\bibitem [{\citenamefont {Pritchard}(1992)}]{pritchard92}%
  \BibitemOpen
  \bibfield  {author} {\bibinfo {author} {\bibfnamefont {W.}~\bibnamefont
  {Pritchard}},\ }\bibfield  {title} {\enquote {\bibinfo {title} {The brain in
  fractal time: 1/f-like power spectrum scaling of the human
  electroencephalogram},}\ }\href@noop {} {\bibfield  {journal} {\bibinfo
  {journal} {Int. Journal of Neurosci.}\ }\textbf {\bibinfo {volume} {66}},\
  \bibinfo {pages} {119--129} (\bibinfo {year} {1992})}\BibitemShut {NoStop}%
\bibitem [{\citenamefont {Zarahn}, \citenamefont {Aguirre},\ and\ \citenamefont
  {Esposito}(1997)}]{zara97}%
  \BibitemOpen
  \bibfield  {author} {\bibinfo {author} {\bibfnamefont {E.}~\bibnamefont
  {Zarahn}}, \bibinfo {author} {\bibfnamefont {G.~K.}\ \bibnamefont {Aguirre}},
  \ and\ \bibinfo {author} {\bibfnamefont {M.~D.}\ \bibnamefont {Esposito}},\
  }\bibfield  {title} {\enquote {\bibinfo {title} {Empirical analysis of bold
  fmri statistics. i. spatially unsmoothed data collected under null-hypothesis
  conditions},}\ }\href@noop {} {\bibfield  {journal} {\bibinfo  {journal}
  {Neuroimage}\ }\textbf {\bibinfo {volume} {5}},\ \bibinfo {pages} {179}
  (\bibinfo {year} {1997})}\BibitemShut {NoStop}%
\bibitem [{\citenamefont {Yamamoto}\ and\ \citenamefont
  {Hughson}(1993)}]{yama93}%
  \BibitemOpen
  \bibfield  {author} {\bibinfo {author} {\bibfnamefont {M.}~\bibnamefont
  {Yamamoto}}\ and\ \bibinfo {author} {\bibfnamefont {R.}~\bibnamefont
  {Hughson}},\ }\bibfield  {title} {\enquote {\bibinfo {title} {Extracting
  fractal components from time series},}\ }\href@noop {} {\bibfield  {journal}
  {\bibinfo  {journal} {Physica D}\ }\textbf {\bibinfo {volume} {68}},\
  \bibinfo {pages} {250--264} (\bibinfo {year} {1993})}\BibitemShut {NoStop}%
\bibitem [{\citenamefont {H\"{a}m\"{a}l\"{a}inen}\ \emph
  {et~al.}(1993)\citenamefont {H\"{a}m\"{a}l\"{a}inen}, \citenamefont {Hari},
  \citenamefont {IImoniemi}, \citenamefont {Knuutila},\ and\ \citenamefont
  {Lounasmaa}}]{hama93}%
  \BibitemOpen
  \bibfield  {author} {\bibinfo {author} {\bibfnamefont {M.}~\bibnamefont
  {H\"{a}m\"{a}l\"{a}inen}}, \bibinfo {author} {\bibfnamefont {R.}~\bibnamefont
  {Hari}}, \bibinfo {author} {\bibfnamefont {R.}~\bibnamefont {IImoniemi}},
  \bibinfo {author} {\bibfnamefont {J.}~\bibnamefont {Knuutila}}, \ and\
  \bibinfo {author} {\bibfnamefont {O.}~\bibnamefont {Lounasmaa}},\ }\bibfield
  {title} {\enquote {\bibinfo {title} {Magnetoencephalography-theory,
  instrumentation, and applications to noninvasive studies of the working human
  brain},}\ }\href@noop {} {\bibfield  {journal} {\bibinfo  {journal} {Rev.
  Mod. Phys.}\ }\textbf {\bibinfo {volume} {65}},\ \bibinfo {pages} {413}
  (\bibinfo {year} {1993})}\BibitemShut {NoStop}%
\bibitem [{\citenamefont {Linkenkaer-Hansen}\ \emph {et~al.}(2001)\citenamefont
  {Linkenkaer-Hansen}, \citenamefont {Nikouline}, \citenamefont {Palva},\ and\
  \citenamefont {IImoniemi}}]{linken01}%
  \BibitemOpen
  \bibfield  {author} {\bibinfo {author} {\bibfnamefont {K.}~\bibnamefont
  {Linkenkaer-Hansen}}, \bibinfo {author} {\bibfnamefont {V.~V.}\ \bibnamefont
  {Nikouline}}, \bibinfo {author} {\bibfnamefont {J.~M.}\ \bibnamefont
  {Palva}}, \ and\ \bibinfo {author} {\bibfnamefont {R.~J.}\ \bibnamefont
  {IImoniemi}},\ }\bibfield  {title} {\enquote {\bibinfo {title} {Long-range
  temporal correlations and scaling behavior in human brain oscillations},}\
  }\href@noop {} {\bibfield  {journal} {\bibinfo  {journal} {Journal of
  Neuroscience}\ }\textbf {\bibinfo {volume} {21}},\ \bibinfo {pages}
  {1370--1377} (\bibinfo {year} {2001})}\BibitemShut {NoStop}%
\bibitem [{\citenamefont {He}\ \emph {et~al.}(2014)\citenamefont {He},
  \citenamefont {Zempel}, \citenamefont {Snyder},\ and\ \citenamefont
  {Raichle}}]{he2014}%
  \BibitemOpen
  \bibfield  {author} {\bibinfo {author} {\bibfnamefont {B.~J.}\ \bibnamefont
  {He}}, \bibinfo {author} {\bibfnamefont {J.~M.}\ \bibnamefont {Zempel}},
  \bibinfo {author} {\bibfnamefont {A.~Z.}\ \bibnamefont {Snyder}}, \ and\
  \bibinfo {author} {\bibfnamefont {M.~E.}\ \bibnamefont {Raichle}},\
  }\bibfield  {title} {\enquote {\bibinfo {title} {The temporal structures and
  functional significance of scale-free brain activity},}\ }\href@noop {}
  {\bibfield  {journal} {\bibinfo  {journal} {Neuron}\ }\textbf {\bibinfo
  {volume} {66}},\ \bibinfo {pages} {353--369} (\bibinfo {year}
  {2014})}\BibitemShut {NoStop}%
\bibitem [{\citenamefont {Tang}, \citenamefont {Bak},\ and\ \citenamefont
  {Wiesenfeld}(1988)}]{btw:soc}%
  \BibitemOpen
  \bibfield  {author} {\bibinfo {author} {\bibfnamefont {C.}~\bibnamefont
  {Tang}}, \bibinfo {author} {\bibfnamefont {P.}~\bibnamefont {Bak}}, \ and\
  \bibinfo {author} {\bibfnamefont {K.}~\bibnamefont {Wiesenfeld}},\ }\bibfield
   {title} {\enquote {\bibinfo {title} {{Self-organized criticality}},}\
  }\href@noop {} {\bibfield  {journal} {\bibinfo  {journal} {Phys. Rev. A}\
  }\textbf {\bibinfo {volume} {38}} (\bibinfo {year} {1988})}\BibitemShut
  {NoStop}%
\bibitem [{\citenamefont {Beggs}\ and\ \citenamefont
  {Plenz}(2003)}]{plenz:pl03}%
  \BibitemOpen
  \bibfield  {author} {\bibinfo {author} {\bibfnamefont {J.~M.}\ \bibnamefont
  {Beggs}}\ and\ \bibinfo {author} {\bibfnamefont {D.}~\bibnamefont {Plenz}},\
  }\bibfield  {title} {\enquote {\bibinfo {title} {{Neuronal avalanches in
  neocortical circuits}},}\ }\href@noop {} {\bibfield  {journal} {\bibinfo
  {journal} {J. Neurosci.}\ }\textbf {\bibinfo {volume} {23}},\ \bibinfo
  {pages} {11167--11177} (\bibinfo {year} {2003})}\BibitemShut {NoStop}%
\bibitem [{\citenamefont {Shriki}\ \emph {et~al.}(2013)\citenamefont {Shriki},
  \citenamefont {Alstott}, \citenamefont {Carver}, \citenamefont {Holroyd},
  \citenamefont {Hanson}, \citenamefont {Smith}, \citenamefont {Coppola},
  \citenamefont {Bullmore},\ and\ \citenamefont {Plenz}}]{shriki13}%
  \BibitemOpen
  \bibfield  {author} {\bibinfo {author} {\bibfnamefont {O.}~\bibnamefont
  {Shriki}}, \bibinfo {author} {\bibfnamefont {J.}~\bibnamefont {Alstott}},
  \bibinfo {author} {\bibfnamefont {F.}~\bibnamefont {Carver}}, \bibinfo
  {author} {\bibfnamefont {T.}~\bibnamefont {Holroyd}}, \bibinfo {author}
  {\bibfnamefont {R.~N.~A.}\ \bibnamefont {Hanson}}, \bibinfo {author}
  {\bibfnamefont {M.~L.}\ \bibnamefont {Smith}}, \bibinfo {author}
  {\bibfnamefont {R.}~\bibnamefont {Coppola}}, \bibinfo {author} {\bibfnamefont
  {E.}~\bibnamefont {Bullmore}}, \ and\ \bibinfo {author} {\bibfnamefont
  {D.}~\bibnamefont {Plenz}},\ }\bibfield  {title} {\enquote {\bibinfo {title}
  {Neuronal avalanches in the resting meg of the human brain},}\ }\href@noop {}
  {\bibfield  {journal} {\bibinfo  {journal} {J. Neurosci.}\ }\textbf {\bibinfo
  {volume} {33}},\ \bibinfo {pages} {7079--7090} (\bibinfo {year}
  {2013})}\BibitemShut {NoStop}%
\bibitem [{\citenamefont {Pruessner}(2012)}]{pruessner}%
  \BibitemOpen
  \bibfield  {author} {\bibinfo {author} {\bibfnamefont {G.}~\bibnamefont
  {Pruessner}},\ }\href@noop {} {\emph {\bibinfo {title} {Self-Organised
  Criticality: Theory, Models and Characterisation}}}\ (\bibinfo  {publisher}
  {Cambridge University Press},\ \bibinfo {year} {2012})\BibitemShut {NoStop}%
\bibitem [{\citenamefont {de~Arcangelis}, \citenamefont {Perrone-Capano},\ and\
  \citenamefont {Herrmann}(2006)}]{lda:06}%
  \BibitemOpen
  \bibfield  {author} {\bibinfo {author} {\bibfnamefont {L.}~\bibnamefont
  {de~Arcangelis}}, \bibinfo {author} {\bibfnamefont {C.}~\bibnamefont
  {Perrone-Capano}}, \ and\ \bibinfo {author} {\bibfnamefont {H.~J.}\
  \bibnamefont {Herrmann}},\ }\bibfield  {title} {\enquote {\bibinfo {title}
  {Self-organized criticality model for brain plasticity},}\ }\href@noop {}
  {\bibfield  {journal} {\bibinfo  {journal} {Phys. Rev. Lett.}\ }\textbf
  {\bibinfo {volume} {96}},\ \bibinfo {pages} {028107} (\bibinfo {year}
  {2006})}\BibitemShut {NoStop}%
\bibitem [{\citenamefont {Lombardi}\ \emph {et~al.}(2012)\citenamefont
  {Lombardi}, \citenamefont {Herrmann}, \citenamefont {Perrone-Capano},
  \citenamefont {Plenz},\ and\ \citenamefont {de~Arcangelis}}]{fl2012}%
  \BibitemOpen
  \bibfield  {author} {\bibinfo {author} {\bibfnamefont {F.}~\bibnamefont
  {Lombardi}}, \bibinfo {author} {\bibfnamefont {H.~J.}\ \bibnamefont
  {Herrmann}}, \bibinfo {author} {\bibfnamefont {C.}~\bibnamefont
  {Perrone-Capano}}, \bibinfo {author} {\bibfnamefont {D.}~\bibnamefont
  {Plenz}}, \ and\ \bibinfo {author} {\bibfnamefont {L.}~\bibnamefont
  {de~Arcangelis}},\ }\bibfield  {title} {\enquote {\bibinfo {title} {Balance
  between excitation and inhibition controls the temporal organization of
  neuronal avalanches},}\ }\href@noop {} {\bibfield  {journal} {\bibinfo
  {journal} {Phys. Rev. Lett}\ }\textbf {\bibinfo {volume} {108}},\ \bibinfo
  {pages} {228703} (\bibinfo {year} {2012})}\BibitemShut {NoStop}%
\bibitem [{\citenamefont {Lombardi}\ \emph {et~al.}(2014)\citenamefont
  {Lombardi}, \citenamefont {Herrmann}, \citenamefont {Plenz},\ and\
  \citenamefont {de~Arcangelis}}]{frontiers}%
  \BibitemOpen
  \bibfield  {author} {\bibinfo {author} {\bibfnamefont {F.}~\bibnamefont
  {Lombardi}}, \bibinfo {author} {\bibfnamefont {H.~J.}\ \bibnamefont
  {Herrmann}}, \bibinfo {author} {\bibfnamefont {D.}~\bibnamefont {Plenz}}, \
  and\ \bibinfo {author} {\bibfnamefont {L.}~\bibnamefont {de~Arcangelis}},\
  }\bibfield  {title} {\enquote {\bibinfo {title} {On the temporal organization
  of neuronal avalanches},}\ }\href@noop {} {\bibfield  {journal} {\bibinfo
  {journal} {Front. Syst. Neurosci.}\ }\textbf {\bibinfo {volume} {8}},\
  \bibinfo {pages} {204. doi:10.3389/fnsys.2014.00204} (\bibinfo {year}
  {2014})}\BibitemShut {NoStop}%
\bibitem [{\citenamefont {Eguiluz}\ \emph {et~al.}(2005)\citenamefont
  {Eguiluz}, \citenamefont {Chialvo}, \citenamefont {Cecchi}, \citenamefont
  {Baliki},\ and\ \citenamefont {Apkarian}}]{chia:sfn}%
  \BibitemOpen
  \bibfield  {author} {\bibinfo {author} {\bibfnamefont {V.~M.}\ \bibnamefont
  {Eguiluz}}, \bibinfo {author} {\bibfnamefont {D.~R.}\ \bibnamefont
  {Chialvo}}, \bibinfo {author} {\bibfnamefont {G.~A.}\ \bibnamefont {Cecchi}},
  \bibinfo {author} {\bibfnamefont {M.}~\bibnamefont {Baliki}}, \ and\ \bibinfo
  {author} {\bibfnamefont {A.~V.}\ \bibnamefont {Apkarian}},\ }\bibfield
  {title} {\enquote {\bibinfo {title} {Scale-free brain functional networks},}\
  }\href@noop {} {\bibfield  {journal} {\bibinfo  {journal} {Phys. Rev. Lett.}\
  }\textbf {\bibinfo {volume} {94}},\ \bibinfo {pages} {018102} (\bibinfo
  {year} {2005})}\BibitemShut {NoStop}%
\bibitem [{\citenamefont {Roerig}\ and\ \citenamefont {Chen}(2002)}]{roerig02}%
  \BibitemOpen
  \bibfield  {author} {\bibinfo {author} {\bibfnamefont {B.}~\bibnamefont
  {Roerig}}\ and\ \bibinfo {author} {\bibfnamefont {B.}~\bibnamefont {Chen}},\
  }\bibfield  {title} {\enquote {\bibinfo {title} {Relationships of local
  inhibitory and excitatory circuits to orientation preference maps in ferret
  visual cortex},}\ }\href@noop {} {\bibfield  {journal} {\bibinfo  {journal}
  {Cereb. Cortex.}\ }\textbf {\bibinfo {volume} {12}},\ \bibinfo {pages}
  {187--198} (\bibinfo {year} {2002})}\BibitemShut {NoStop}%
\bibitem [{\citenamefont {Bonifazi}\ \emph {et~al.}(2009)\citenamefont
  {Bonifazi}, \citenamefont {Goldin}, \citenamefont {Picardo}, \citenamefont
  {Jorquera}, \citenamefont {Cattani}, \citenamefont {Bianconi}, \citenamefont
  {Represa}, \citenamefont {Ben-Ari},\ and\ \citenamefont
  {Cossart}}]{bonifaz09}%
  \BibitemOpen
  \bibfield  {author} {\bibinfo {author} {\bibfnamefont {P.}~\bibnamefont
  {Bonifazi}}, \bibinfo {author} {\bibfnamefont {M.}~\bibnamefont {Goldin}},
  \bibinfo {author} {\bibfnamefont {M.~A.}\ \bibnamefont {Picardo}}, \bibinfo
  {author} {\bibfnamefont {I.}~\bibnamefont {Jorquera}}, \bibinfo {author}
  {\bibfnamefont {A.}~\bibnamefont {Cattani}}, \bibinfo {author} {\bibfnamefont
  {G.}~\bibnamefont {Bianconi}}, \bibinfo {author} {\bibfnamefont
  {A.}~\bibnamefont {Represa}}, \bibinfo {author} {\bibfnamefont
  {Y.}~\bibnamefont {Ben-Ari}}, \ and\ \bibinfo {author} {\bibfnamefont
  {R.}~\bibnamefont {Cossart}},\ }\bibfield  {title} {\enquote {\bibinfo
  {title} {Gabaergic hub neurons orchestrate synchrony in developing
  hippocampal networks},}\ }\href@noop {} {\bibfield  {journal} {\bibinfo
  {journal} {Science}\ }\textbf {\bibinfo {volume} {326}} (\bibinfo {year}
  {2009})}\BibitemShut {NoStop}%
\bibitem [{Note1()}]{Note1}%
  \BibitemOpen
  \bibinfo {note} {Given two neurons, $i$ and $j$, and a synaptic connection
  directed from $i$ to $j$, $i$ is called presynaptic and $j$ postsynaptic
  neuron.}\BibitemShut {Stop}%
\bibitem [{\citenamefont {Rios}\ and\ \citenamefont {Zhang}(1999)}]{rios:99}%
  \BibitemOpen
  \bibfield  {author} {\bibinfo {author} {\bibfnamefont {P.~D.~L.}\
  \bibnamefont {Rios}}\ and\ \bibinfo {author} {\bibfnamefont {Y.-C.}\
  \bibnamefont {Zhang}},\ }\bibfield  {title} {\enquote {\bibinfo {title}
  {Universal 1/f noise from dissipative self-organized criticality model},}\
  }\href@noop {} {\bibfield  {journal} {\bibinfo  {journal} {Phys.Rev.Lett.}\
  }\textbf {\bibinfo {volume} {82}},\ \bibinfo {pages} {472} (\bibinfo {year}
  {1999})}\BibitemShut {NoStop}%
\bibitem [{\citenamefont {Poil}, \citenamefont {van Ooyen},\ and\ \citenamefont
  {Linkenkaer-Hansen}(2008)}]{poil08}%
  \BibitemOpen
  \bibfield  {author} {\bibinfo {author} {\bibfnamefont {S.-S.}\ \bibnamefont
  {Poil}}, \bibinfo {author} {\bibfnamefont {A.}~\bibnamefont {van Ooyen}}, \
  and\ \bibinfo {author} {\bibfnamefont {K.}~\bibnamefont
  {Linkenkaer-Hansen}},\ }\bibfield  {title} {\enquote {\bibinfo {title}
  {Avalanche dynamics of human brain oscillations: Relation to critical
  branching processes and temporal correlations},}\ }\href@noop {} {\bibfield
  {journal} {\bibinfo  {journal} {Human Brain Mapping}\ }\textbf {\bibinfo
  {volume} {29}},\ \bibinfo {pages} {770--777} (\bibinfo {year}
  {2008})}\BibitemShut {NoStop}%
\bibitem [{\citenamefont {Jensen}(1998)}]{jsn:soc}%
  \BibitemOpen
  \bibfield  {author} {\bibinfo {author} {\bibfnamefont {H.~J.}\ \bibnamefont
  {Jensen}},\ }\href@noop {} {\emph {\bibinfo {title} {Self-Organized
  Criticality}}}\ (\bibinfo  {publisher} {Cambridge University Press},\
  \bibinfo {year} {1998})\BibitemShut {NoStop}%
\bibitem [{\citenamefont {Kuntz}\ and\ \citenamefont
  {Sethna}(2000)}]{sethna00}%
  \BibitemOpen
  \bibfield  {author} {\bibinfo {author} {\bibfnamefont {M.}~\bibnamefont
  {Kuntz}}\ and\ \bibinfo {author} {\bibfnamefont {J.}~\bibnamefont {Sethna}},\
  }\bibfield  {title} {\enquote {\bibinfo {title} {Noise in disordered systems:
  The power spectrum and dynamic exponents in avalanche models},}\ }\href@noop
  {} {\bibfield  {journal} {\bibinfo  {journal} {Phys. Rev. B}\ }\textbf
  {\bibinfo {volume} {62}} (\bibinfo {year} {2000})}\BibitemShut {NoStop}%
\bibitem [{\citenamefont {Davidsen}\ and\ \citenamefont
  {Paczuski}(2002)}]{davidsen02}%
  \BibitemOpen
  \bibfield  {author} {\bibinfo {author} {\bibfnamefont {J.}~\bibnamefont
  {Davidsen}}\ and\ \bibinfo {author} {\bibfnamefont {M.}~\bibnamefont
  {Paczuski}},\ }\bibfield  {title} {\enquote {\bibinfo {title} {$1/f^{\alpha}$
  noise from correlations between avalanches in self-organized criticality},}\
  }\href@noop {} {\bibfield  {journal} {\bibinfo  {journal} {Phys.Rev.E}\
  }\textbf {\bibinfo {volume} {66}},\ \bibinfo {pages} {050101} (\bibinfo
  {year} {2002})}\BibitemShut {NoStop}%
\bibitem [{\citenamefont {M.}\ and\ \citenamefont {C.}(2006)}]{baie:sand}%
  \BibitemOpen
  \bibfield  {author} {\bibinfo {author} {\bibfnamefont {B.}~\bibnamefont
  {M.}}\ and\ \bibinfo {author} {\bibfnamefont {M.}~\bibnamefont {C.}},\
  }\bibfield  {title} {\enquote {\bibinfo {title} {Realistic time correlations
  in sandpiles},}\ }\href@noop {} {\bibfield  {journal} {\bibinfo  {journal}
  {Europhys. Lett.}\ }\textbf {\bibinfo {volume} {75}} (\bibinfo {year}
  {2006})}\BibitemShut {NoStop}%
\bibitem [{\citenamefont {Shew}\ \emph {et~al.}(2009)\citenamefont {Shew},
  \citenamefont {Yang}, \citenamefont {Petermann}, \citenamefont {Roy},\ and\
  \citenamefont {Plenz}}]{shew09}%
  \BibitemOpen
  \bibfield  {author} {\bibinfo {author} {\bibfnamefont {W.}~\bibnamefont
  {Shew}}, \bibinfo {author} {\bibfnamefont {H.}~\bibnamefont {Yang}}, \bibinfo
  {author} {\bibfnamefont {T.}~\bibnamefont {Petermann}}, \bibinfo {author}
  {\bibfnamefont {R.}~\bibnamefont {Roy}}, \ and\ \bibinfo {author}
  {\bibfnamefont {D.}~\bibnamefont {Plenz}},\ }\bibfield  {title} {\enquote
  {\bibinfo {title} {Neuronal avalanches imply maximum dynamic range in
  cortical networks at criticality},}\ }\href@noop {} {\bibfield  {journal}
  {\bibinfo  {journal} {J.Neurosci.}\ }\textbf {\bibinfo {volume} {29}},\
  \bibinfo {pages} {15595--15600} (\bibinfo {year} {2009})}\BibitemShut
  {NoStop}%
\bibitem [{\citenamefont {Kinouchi}\ and\ \citenamefont
  {Copelli}(2006)}]{copelli06}%
  \BibitemOpen
  \bibfield  {author} {\bibinfo {author} {\bibfnamefont {O.}~\bibnamefont
  {Kinouchi}}\ and\ \bibinfo {author} {\bibfnamefont {M.}~\bibnamefont
  {Copelli}},\ }\bibfield  {title} {\enquote {\bibinfo {title} {Optimal
  dynamical range of excitable networks at criticality},}\ }\href@noop {}
  {\bibfield  {journal} {\bibinfo  {journal} {Nat. Phys.}\ }\textbf {\bibinfo
  {volume} {2}},\ \bibinfo {pages} {348} (\bibinfo {year} {2006})}\BibitemShut
  {NoStop}%
\bibitem [{\citenamefont {Lombardi}\ \emph {et~al.}(2016)\citenamefont
  {Lombardi}, \citenamefont {Herrmann}, \citenamefont {Plenz},\ and\
  \citenamefont {de~Arcangelis}}]{fl2016}%
  \BibitemOpen
  \bibfield  {author} {\bibinfo {author} {\bibfnamefont {F.}~\bibnamefont
  {Lombardi}}, \bibinfo {author} {\bibfnamefont {H.~J.}\ \bibnamefont
  {Herrmann}}, \bibinfo {author} {\bibfnamefont {D.}~\bibnamefont {Plenz}}, \
  and\ \bibinfo {author} {\bibfnamefont {L.}~\bibnamefont {de~Arcangelis}},\
  }\bibfield  {title} {\enquote {\bibinfo {title} {Temporal correlations in
  avalanche occurrence},}\ }\href@noop {} {\bibfield  {journal} {\bibinfo
  {journal} {Sci. Rep.}\ }\textbf {\bibinfo {volume} {6}},\ \bibinfo {pages}
  {24690; doi: 10.1038/srep24690} (\bibinfo {year} {2016})}\BibitemShut
  {NoStop}%
\end{thebibliography}%

\end{document}